\documentstyle[12pt]{article}

\begin{document}
{\sf \begin{center} \noindent {\Large\bf Geodesic dynamo chaotic
flows and non-Anosov maps in twisted magnetic flux
tubes}\\[2mm]

by \\[0.1cm]

{\sl  L.C. Garcia de Andrade}\\Departamento de F\'{\i}sica
Te\'orica -- IF -- Universidade do Estado do Rio de Janeiro-UERJ\\[-3mm]
Rua S\~ao Francisco Xavier, 524\\[-3mm]
Cep 20550-003, Maracan\~a, Rio de Janeiro, RJ, Brasil\\[-3mm]
Electronic mail address: garcia@dft.if.uerj.br\\[-3mm]
\vspace{1cm} {\bf Abstract}
\end{center}
\paragraph*{}
Recently Tang and Boozer [{\textbf{Phys. Plasmas (2000)}}], have
investigated the anisotropies in magnetic field dynamo evolution,
from local Lyapunov exponents, giving rise to a metric tensor, in
the Alfven twist in magnetic flux tubes (MFTs). Thiffeault and
Boozer [\textbf{Chaos}(2001)] have investigated the how the
vanishing of Riemann curvature constrained the Lyapunov exponential
stretching of chaotic flows. In this paper, Tang-Boozer-Thiffeault
differential geometric framework is used to investigate effects of
twisted magnetic flux tube filled with helical chaotic flows on the
Riemann curvature tensor. When Frenet torsion is positive, the
Riemann curvature is unstable, while the negative torsion induces an
stability  when time $t\rightarrow{\infty}$. This enhances the
dynamo action inside the MFTs. The Riemann metric, depends on the
radial random flows along the poloidal and toroidal directions. The
Anosov flows has been applied by Arnold, Zeldovich, Ruzmaikin and
Sokoloff [\textbf{JETP (1982)}] to build a uniformly stretched
dynamo flow solution, based on Arnold's Cat Map. It is easy to show
that when the random radial flow vanishes, the magnetic field
vanishes, since the exponential Lyapunov stretches vanishes. This is
an example of the application of the Vishik's anti-fast dynamo
theorem in the magnetic flux tubes. Geodesic flows of both Arnold
and twisted MFT dynamos are investigated. It is shown that a
constant random radial flow can be obtained from the geodesic
equation. Throughout the paper one assumes, the reasonable plasma
astrophysical hypothesis of the weak torsion. Pseudo-Anosov dynamo
flows and maps have also been addressed by Gilbert [\textbf{Proc Roy
Soc A London (1993)}].

{\bf PACS numbers:\hfill\parbox[t]{13.5cm}{02.40.Hw:differential
geometries.91.25.Cw-dynamo theories.}}}
\newline

 \section{Introduction}
 To talk about the importance of the developement of the
 concepts of Riemannian geometry \cite{1} to physics, would be enough
 to mention the advances in Einstein's
 gravitational theory, more well known as the general theory of relativity \cite{2}. More recently, uses in other parts of physics such as string theory of defects and
 solid state theory \cite{3} have also experienced the same success. Yet more recently, Thiffeault , Tang and Boozer, on a series of papers \cite{4,5,6,7} have
 investigate the constraints experienced by chaotic flows, when the Riemann curvature tensor vanishes. In this paper, a slightly more general application
 of these ideas are applied to helical chaotic flows inside the twisted magnetic flux tubes
 \cite{8,9} is given. The role of random radial flows in Lyapunov exponential stretching, so fundamental to the kinematic dynamo problem, is investigated
 throughout the paper. The investigation of Anosov diffeomorpisms \cite{9}, an important mathematical tool from the theory of
 dynamical systems, has been often used, in connection with
 the investigation of dynamo flows and maps \cite{9} such as the Arnold's Cat Map \cite{9} on the torus, useful in mixing \cite{10} problems in the physics of
 fluids \cite{4}. One of the main properties of
 the Anosov maps is that they yield Lyapunov exponential of the
 chaotic exponential stretching, which are constant everywhere \cite{5}. In their paper,
 Thiffeault and Boozer \cite{5}, obtained geometrical constraints in
 two and three dimensions, with coordinates $\textbf{x}\in\textbf{R}^{2}$ and
 $\textbf{x}'\in\textbf{R}^{3}$. Consequences of the vanishing of the
 Riemann curvature tensors, or Riemann-flat conditions in the mathematicians jargon,
 includes constraining the modification of the growth rate of magnetic fields
 in the kinematic dynamo problem or the mixing properties of chaotic
 advection-diffusive equation. They conclude that, the Lyapunov exponents are locally small in regions where, Riemann curvature of the stable manifold,
 is large. Though Euclidean metrics were used in their work to allow
 for the vanishing of the Riemann curvature, here this hypothesis is
 relaxed, which still allows one to obtain similar results for the
 flux tube endowed with chaotic helical flows. The Riemann and Ricci
 curvatures are computed in terms of the Lyapunov exponential
 stretching and establishing the conditions on the Frenet torsion of
 the tube axis to obtain positive Lyapunov exponents which leads to
 the fast dynamo action for chaotic flows permeatting the tube.
 \newline
    This last condition is fundamental for dynamo action \cite{8}.
 Random radial flows \cite{10,11} are shown to be fundamental, for the chaotic flows to experience a non-vanishing exponential stretching. Since these MFTs
 are fundamental in astrophysical plasmas and solar chaotic plasmas
 in tokamaks and stellarators, as well in the Perm torus of liquid sodium dynamo \cite{12}, the better understanding of chaotic flow, seems to be useful to
 future developement of plasma physics and dynamo theory. Therefore a good
 understanding of the behaviour of these rotating torus flows is of
 utmost importance, in future dynamo experiments. This paper is organised as follows. In section 2, special types of chaotic flows in twisted MFTs are presented such as the
 torus automorphisms, called Arnold's Cat Map, which led Arnold \cite{9} to build a uniform stretching dynamo solution of self-induction equation. In this case though the
 Jacobian differs from the Arnold, Anosov map one, the same eigenvalues are obtained, and therefore the dynamo chaotic flow is stretched along one direction
 and compressed along the other. Section 3 generalizes the Anosov flows to the case where the stretch along toroidal direction is not constant and therefore the
 Lyapunov metric of the tube leads to the computation of Riemann and Ricci curvature in the twisted MFTs metric. This kind of twisted geometry is then used to established constraints
 along the section to established the stability relation to the torsion of the tube, which in turn is proportional to the
 twist. Section 4 presents the discussion and conclusions.
 \newpage
\section{Hyperbolic dynamo maps and flows in MFTs}
\vspace{1cm} This section contains a brief review and some new
material on hyperbolic flows and dynamo maps. While is not the main
section of the paper, it is supposed to pave the way to the next
section where the main results are given. Let us start by defining
what exactly is an hyperbolic map. Anosov maps are actually an
example of hyperbolic map with extra constraints which one shall not
bother here. An hyperbolic map on a torus may be defined as an
automorphism where the determinant of its Jacobian is equal to one.
This constraint shall be used in this section, in analogy with
Thiffeault and Boozer \cite{4} constraints to place constraints on
the dynamos and Riemannian geometry of the tubes. Let us consider
the following two-dimensional twist map \cite{13} Jacobian matrix
\begin{equation}
\textbf{J}_{\textbf{twist}}=\pmatrix{1&-{\tau}_{0}\cr0&K_{0}\cr}\qquad
\label{1}
\end{equation}
Here $K_{0}$ is the toroidal stretching along the tube metric
\cite{14} given by
\begin{equation}
dl^{2}= dr^{2}+r^{2}d{{\theta}_{R}}^{2}+K^{2}(r,s)ds^{2} \label{2}
\end{equation}
where of course here, the internal cross-section radius $r$ is
constant for the two dimensional Jacobian and the twist
transformation is
\begin{equation}
{\theta}(s):={\theta}_{R}-\int{{\tau}(s)ds} \label{3}
\end{equation}
where the integral in this expression is the total torsion. Here,
since one is considering helical chaotic flows, the Frenet torsion
and curvature of the tube axis are constants and fixed to be equal.
As one shall note in the next section the introduction of coordinate
r into the stretching $K(r,s)=1-r{\kappa}(s)cos{\theta}$ allows one
to has Riemann and Ricci curvature that do not vanish identically.
Of course here, one have considered that $K_{0}$ is a constant or
uniform stretching so important in chaos \cite{14}. A small lemma
can be proved here by constrained the Jacobian transformation to be
hyperbolic. Thus \newline \textbf{lemma}: Hyperbolic maps with
constant stretch in the twist MFTs, leads necessarilly to a thin
tube. \textbf{Proof}: This comes immeadiatly from the determinant of
matrix (\ref{1}) equals to one
\begin{equation}
Det[\textbf{J}_{\textbf{twist}}]=K_{0}=1\label{4}
\end{equation}
which by examining the expression for $K(r,s)$ shows that $K(0,s)=1$
is the thin tube approximation, as one wishes to prove.Let us now
compute the eigenvalue problem. Just to show that the twist tube
Jacobian above is a general case of the torus automorphism by
Arnold, let us consider a special transformation where the twist is
fixed while the radius is contracted to half his size and the
stretch is double, similar to what happens in the stretch-twist-fold
(STF) dynamo mechanism proposed by Vainshtein and Zeldovich
\cite{15}. Mathematically this transformation is
\begin{equation}
r'=\frac{1}{2}r \label{5}
\end{equation}
\begin{equation}
s'={2}s \label{6}
\end{equation}
while ${\theta}'={\theta}_{R}$. Besides the torsion ${\tau}_{0}$ is
normalized to one, since one does not want to deal with untwisted
tubes. Therefore the three-dimensional Jacobian
\begin{equation}
\textbf{J}_{\textbf{3D}}=\pmatrix{K_{1}&0&0\cr0&1&-{\tau}_{0}\cr0&0&K_{0}\cr}\qquad
\label{7}
\end{equation}
where $K_{1}$ is the stretching (contraction) of the radial
direction, takes the form
\begin{equation}
\textbf{J}_{\textbf{3D}}=\pmatrix{\frac{1}{2}&0&0\cr0&1&-1\cr0&0&2\cr}\qquad
\label{8}
\end{equation}
By taking the determinant of this Jacobian one obtains it is equal
to one, or hyperbolic, and besides the eigenvalue problem
\begin{equation}
Det[\textbf{J}_{\textbf{3D}}-{\lambda}\textbf{1}]=0 \label{9}
\end{equation}
where $\textbf{1}$ is the unity matrix
\begin{equation}
\textbf{1}=\pmatrix{{1}&0&0\cr0&1&0\cr0&0&1\cr}\qquad \label{10}
\end{equation}
${\lambda}$ being the eigenvalue. The determinant leads to the
following second-order algebraic equation
\begin{equation}
{\lambda}^{2}-3{\lambda}+1=0\label{11}
\end{equation}
Note that this equation leads to the Arnold eigenvalues
\begin{equation}
{\lambda}_{\pm}=\frac{3\pm{\sqrt{5}}}{2}\label{12}
\end{equation}
where ${\lambda}_{+}>0$ represents the stretching, while
${\lambda}_{-}<0$ is the contraction of the flow. In the Arnold's
case \cite{13} the stretching is uniform. In the next section, one
shall be concerned with the non-uniform stretching where the
stretching Riemann metric factors, do depend upon the coordinates.
Another interesting lemma can be proved for the three-dimensional
matrix, as
\newline
\textbf{lemma}:\newline The hyperbolicity condition applied to twist
map in ${\textbf{R}^{3}}$, given by the above Jacobian, implies that
the magnetic flux tubes stretching and contraction obeys the
following conditions: $K_{0}={K_{1}}^{-1}$, which tell us that the
stretching is the inverse of contraction as foreseen by our physical
intuition.
\newline
\textbf{Proof}:\newline The proof is trivial, and obtained by
applying the $Det{\textbf{J}}=1$ condition to the three-dimensional
Jacobian $\textbf{J}_{\textbf{3D}}$ above. This yields that
$K_{0}K_{1}=1$ as we wish to prove. \newline The domain of the flow
is given by a compact Riemannian manifold represented as the product
of a torus $\textbf{T}^{2}$ and the closed interval $[0,1]$ of
$0\le{z}\le{1}$ \cite{9}. This results in the Arnold Riemann metric
\begin{equation}
ds^{2}=e^{-{\lambda}z}dp^{2}+e^{{\lambda}z}dq^{2}+dz^{2} \label{13}
\end{equation}
which represents the stretching and contraction in distinct
Euclidean directions p and q , respectively induced by the
eigenvalues ${\lambda}_{\pm}=\frac{3\pm{\sqrt{5}}}{2}$ also
corresponding to magnetic eigenvectors. The stationary dynamo flow
considered by Arnold was given by the simple uniformly stretching
flow $\textbf{v}=(0,0,v)$ in $(p,q,z)$ coordinates. More recent
attempts to build a dynamo action by making use of compact
Riemannian geometry includes the case of the fast dynamo of Chiconne
and Latushkin \cite{16}, and the conformally stretched fast dynamo
by Garcia de Andrade \cite{17}. Anosov map in Arnold's cat map, is
given by the cat dynamo Jacobian matrix \cite{4}
\begin{equation}
\textbf{J}_{\textbf{cat}}=\pmatrix{2&1\cr1&1\cr}\qquad \label{14}
\end{equation}
Note that this Jacobian is distinct to the previous Jacobians but
satisfies the hyperbolicity condition and leads to the same
eigenvalues as before. Let us now compute the Riemann curvature of
the Arnold metric
\begin{equation}
R_{1212}={\lambda}^{2} \label{15}
\end{equation}
\begin{equation}
R_{1313}=-{\lambda}^{2} \label{16}
\end{equation}
\begin{equation}
R_{2323}=-{{\lambda}^{2}} \label{17}
\end{equation}
for the Riemann curvature tensor, while the Ricci tensor components
are
\begin{equation}
R_{33}=R=2{\lambda}^{2} \label{18}
\end{equation}
the determinant of the metric $g=1$, implies that the flow is
incompressible as considered previously by Arnold. Note as well,
that, as usual in Anosov systems the Riemann curvature is constant.
\section{Riemann curvature from Lyapunov exponents
and dynamo action}\vspace{1cm} In this section it is shown that in
regions of weak torsion the Riemann and Ricci curvatures acquire
simple forms, in terms of of exponential stretching and torsion.
Negative torsion in helical chaotic flows, leads to singularities in
curvatures in the $t\rightarrow{\infty}$ time limit. A simple
examination of the Jacobians considered in the last section allows
us to see that the Riemann metric of the flux tubes vanish for a
constant stretch-twist and contraction. However since curvature or
folding is a fundamental process in the STF dynamo mechanism, in
this section it is shown that the consideration of a more general
Jacobian makes the manifold acquire a curved Riemannian metric
distinct from the Euclidean norm metrics considered so far. So, for
a non-uniform stretching \cite{14} usually found in general chaotic
flows the Riemann and Ricci curvatures do not vanish identically.
Let us start computing the curvatures corresponding to the general
Riemann metric of the twisted MFT above. With the aid of an
adaptation of the tensor package the computation of the curvatures
is neither tedious nor long and results in
\begin{equation}
R_{1313}=\frac{1}{4}\frac{{{\tau}_{0}}^{2}cos^{2}({\theta})}{(1-{\tau}_{0}rcos{\theta})}
\label{19}
\end{equation}
\begin{equation}
R_{1323}=-\frac{1}{8}\frac{{{\tau}_{0}}^{2}rsin{2{\theta}}}{(1-{\tau}_{0}rcos{\theta})}
\label{20}
\end{equation}
\begin{equation}
R_{2323}=\frac{1}{4}\frac{{{\tau}_{0}}^{2}r^{2}sin{2{\theta}}}{(1-{\tau}_{0}rcos{\theta})}
\label{21}
\end{equation}
for the Riemann curvature tensor, while the Ricci tensor components
are
\begin{equation}
R_{11}=-\frac{1}{4}\frac{{{\tau}_{0}}^{2}cos^{2}{\theta}}{(1-2{\tau}_{0}rcos{\theta}+{{\tau}_{0}}^{2}rcos{\theta})}
\label{22}
\end{equation}
\begin{equation}
R_{12}=-\frac{1}{8}\frac{{{\tau}_{0}}^{2}rsin{2{\theta}}}{(1-2{\tau}_{0}rcos{\theta}+{{\tau}_{0}}^{2}rcos{\theta})}
\label{23}
\end{equation}
\begin{equation}
R_{33}=-\frac{1}{4}\frac{{{\tau}_{0}}^{2}}{(1-{\tau}_{0}rcos{\theta})}
\label{24}
\end{equation}
\begin{equation}
R_{22}=-\frac{1}{8}\frac{{{\tau}_{0}}^{2}rsin{2{\theta}}}{(1-2{\tau}_{0}rcos{\theta}+{{\tau}_{0}}^{2}rcos{\theta})}
\label{25}
\end{equation}
Most of these curvatures can be easily simplified in the weak
torsion case. Let us now give a more dynamical character to these
expressions by considering the Thiffeault-Boozer Riemann metric
relation with the Lyapunov exponents of exponential stretching by
expressing this metric in terms of the directions $\textbf{t}$,
$\textbf{e}_{r}$ and $\textbf{e}_{\theta}$ along the curved tube as
\begin{equation}
g_{ij}={\Lambda}_{r}\textbf{e}_{r}\textbf{e}_{r}+{\Lambda}_{\theta}\textbf{e}_{\theta}\textbf{e}_{\theta}+{\Lambda}_{s}\textbf{t}\textbf{t}
\label{26}
\end{equation}
where $({i,j=r,{\theta},s})$ and ${\Lambda}_{i}$ are the Lyapunov
numbers which are all positive or null. The Lyapunov exponents are
given by
\begin{equation}
{{\lambda}^{\infty}}_{i}=lim_{t\rightarrow\infty}(\frac{ln{\Lambda}_{i}}{2t})
\label{27}
\end{equation}
The infinite symbol over the Lyapunov exponent indicates that this
is a true Lyapunov exponent obtained as the limit of the finite-time
Lyapunov exponent ${\lambda}_{i}$. Let us now compute the values of
the Lyapunov exponents which are fundamental for stretching and
dynamos, in terms of a random radial flow
\begin{equation}
{\lambda}_{i}=lim_{t\rightarrow\infty}(\frac{ln{\Lambda}_{i}}{2t})
\label{28}
\end{equation}
\begin{equation}
<r>=\int{<v_{r}>dt} \label{29}
\end{equation}
where here one shall consider that on a finite-time period, the
random flow can be considered as approximately constant which
reduces the last expression to
\begin{equation}
<r>=<v_{r}>t \label{30}
\end{equation}
Let us now compute the Lyapunov exponents of the curved Riemannian
flux tubes in terms of the random flows as
\begin{equation}
{\lambda}_{r}=lim_{t\rightarrow\infty}(\frac{ln{1}}{2t})=0
\label{31}
\end{equation}
\begin{equation}
{\lambda}_{\theta}=lim_{t\rightarrow\infty}(\frac{r}{t})=<v_{r}>
\label{32}
\end{equation}
and finally
\begin{equation}
{\lambda}_{s}=lim_{t\rightarrow\infty}(\frac{lnK(r,s)}{2t})\approx{lim_{t\rightarrow\infty}(\frac{-{\tau}_{0}rsin{\theta}}{2t})}\approx{-{\tau}_{0}
<v_{r}>sin{\theta}} \label{33}
\end{equation}
By following the computations by Tang and Boozer \cite{5} on the
magnetic fields by the self-induction equation one can say that the
corresponding magnetic field components maybe written in terms of
the Lyapunov exponents as
\begin{equation}
{B}_{\theta}\approx{e^{2{\lambda}_{\theta}t}}=e^{<v_{r}>t}
\label{34}
\end{equation}
\begin{equation}
B_{s}\approx{e^{{\lambda}_{s}t}}=e^{-{\tau}_{0}<v_{r}>sin{\theta}t}
\label{35}
\end{equation}
while the radial magnetic field does not depend on time , which
allows us to say that this is due to the confinement of the radial
flow on the tube, actually in the force-free case considered by
Ricca \cite{18} the flux tube radial magnetic component $B_{r}$ is
assumed to vanish. From the last two expressions it is easy to
observe that the magnetic field is stationary in the absence of
random radial flows, and there is no fast dynamo action as well as
no stretching due to the vanishing of the Lyapunov exponents. This
phenomenon is actually the Vishik's anti-fast dynamo theorem
\cite{19} where no fast-dynamo action can be obtained, in
non-stretching flows. Fast dynamo is obtained when the random flow
has a positive average velocity. Note that in this case at least the
$B_{s}$ toroidal field grows in time, while the poloidal field is
spatially periodic and may even grow in time in the case torsion is
negative. To simplify matters let us now compute the Riemann and
Ricci curvatures for the chaotic flow tube metric
\begin{equation}
d{s_{0}}^{2}=dr^{2}+e^{<v_{r}>t}d{\theta}^{2}+e^{-{\tau}_{0}<v_{r}>cos{\theta}t}ds^{2}
\label{36}
\end{equation}
Note that for this expression a positive torsion shows that one of
the Ljapunov exponents in this Riemann metric is positive
(stretching) while the other is negative, representing contraction.
The only non-vanishing component for the Riemann tensor is
\begin{equation}
R_{2323}=-\frac{1}{4}e^{2<v_{r}>{\tau}_{0}(1-cos{\theta})t}sin^{2}{\theta}\label{37}
\end{equation}
which shows that the Riemann curvature is unstable in the infinite
time limit, if the torsion is positive, while it is stable if the
torsion is negative. The Ricci tensor components are
\begin{equation}
R_{33}=e^{(1-{\tau}_{0}cos{\theta})<v_{r}>t}R_{22}\label{38}
\end{equation}
\begin{equation}
R_{22}=\frac{1}{2}cos{\theta}e^{-(1-{\tau}_{0}<v_{r}>t)}\label{39}
\end{equation}
The Ricci tensor is
\begin{equation}
R=\frac{1}{2}e^{-{\lambda}_{\theta}(1-{\tau}_{0})t}sin^{2}{\theta}
\label{40}
\end{equation}
In all these computations one has assumed the weak torsion
approximation, which is very reasoble for example, in astrophysical
plasmas. Note from the Riemann curvature expression that the Riemann
space is flat when the random radial flow vanishes, and when the
torsion ${\tau}_{0}>1$ the scalar curvature R is singular, or
unstable, when $t\rightarrow{\infty}$. However, since one assumes
here that the weak torsion approximation, the curvature computations
are ingeneral unstable. Since the twist of plasma MFTs is
proportional to torsion of the MFT axis, this is a very reasonable
approximation since the twist in kink solar loops for examples is
very weak of the order of $Tw\approx{10^{-10}} cm^{-1}$. When
torsion vanishes the stationary Riemann curvature is periodic and
reduces to
\begin{equation}
R_{2323}=-\frac{1}{4}sin^{2}{\theta}\label{41}
\end{equation}
Thus when the radial random flow vanishes the Riemann curvature can
be spatially periodically. Let us now compute the determinant of the
Riemann metric
\begin{equation}
g=Det{(g_{ij})}=r^{2}(1-{\tau}_{0}rcos{\theta})^{2}\label{42}
\end{equation}
This might be $g=1$ in the case of incompressible chaotic flow as
${\nabla}.\textbf{v}=0$, which allows us to say that the chaotic
helical flow inside the tube has to be compressible. Tang and Boozer
\cite{5}, have also examine the problem of compressible chaotic
flows. As a final observation one notes that the Frenet curvature of
material lines inside the tube, increases as the Ljapunov exponents
or metric decreases. This can be seen by a simple examination of the
metric factor $g_{ss}=(1-{\kappa}(s)rcos{\theta})^{2}$ since the
Frenet curvature ${\kappa}(s)$ growth implies that this stretching
factor decreases, of course along the Ljapunov exponents since they
depend upon the Riemann metric.
\section{Geodesic flows in chaotic dynamos}
Anosov \cite{14} demonstrated that hyperbolic systems which are
geodesic flows \cite{20}, are necessarilly Anosov. In this section
one shows that as a consequence of the geodesic flow condition
imposed the MFT twisted flows are non-Anosov since their Lyapunov
exponents are not all constants. In this section two given, the
first is the Anosov geodesic flow of Arnold dynamo, while the second
addresses the example of geodesic flows on the twisted MFTs above.
In the first case, the Arnold metric of previous section, yields the
Riemann-Christoffel symbols as
\begin{equation}
{{\Gamma}^{i}}_{jk}=\frac{1}{2}g^{il}(g_{lj,k}+g_{lk,j}-g_{jk,l})\label{43}
\end{equation}
whose components are
\begin{equation}
{{\Gamma}^{1}}_{13}=-\lambda=-{{\Gamma}^{2}}_{23}\label{44}
\end{equation}
\begin{equation}
{{\Gamma}^{3}}_{11}={\lambda}e^{-2{\lambda}z}\label{45}
\end{equation}
\begin{equation}
{{\Gamma}^{3}}_{22}=-{\lambda}e^{2{\lambda}z}\label{46}
\end{equation}
Substitution of these symbols into the geodesic equation
\begin{equation}
\frac{dv^{i}}{dt}+{{\Gamma}^{i}}_{jk}v^{j}v^{k}=0\label{47}
\end{equation}
yields the following equations for the geodesic flow $\textbf{v}$ as
\begin{equation}
\frac{dv^{1}}{dt}+{{\Gamma}^{1}}_{13}v^{1}v^{3}=0\label{48}
\end{equation}
which yields
\begin{equation}
\frac{dv^{1}}{dt}-{\lambda}v^{1}v^{3}=0\label{49}
\end{equation}
Proceeding the same way with the remaining equations yields
\begin{equation}
\frac{dv^{2}}{dt}+{{\lambda}}v^{2}v^{3}=0\label{50}
\end{equation}
and
\begin{equation}
\frac{dv^{3}}{dt}-{\lambda}[e^{-2{\lambda}z}(v^{1})^{2}-e^{2{\lambda}z}(v^{2})^{2}]=0\label{51}
\end{equation}
By assuming that the $v^{3}=constant={v^{3}}_{0}$ one obtains
\begin{equation}
v^{1}=v^{0}e^{{\lambda}[z+{v^{3}}_{0}t]} \label{52}
\end{equation}
\begin{equation}
v^{2}=v^{0}e^{-{\lambda}[z-{v^{3}}_{0}t]}\label{53}
\end{equation}
Since $v^{1}=v^{p}=\frac{dp}{dt}$ and $v^{2}=v^{q}=\frac{dq}{dt}$,
one is able to integrate the geodesic equations to obtain the flow
topology as
\begin{equation}
p^{2}+q^{2}=R^{2}e^{{\lambda}_{L}t}sinh{{\lambda}z}\label{54}
\end{equation}
where ${\lambda}_{L}:=v_{0}{\lambda}$ is the Lyapunov exponent for
the Arnold metric, and $z:=v_{0}t$. This geometry represents a
circle expanding on time and exponentially stretched by the action
of the Lyapunov exponent. Note that the dynamo flow velocity here,
is distinct from the Arnold's one, since the Arnold's dynamo flow
does not possesses components $v^{p}$ and $v^{q}$ as in the example
considered in this section. Let us now proceed computing the same
geodesic flow in the case of twisted MFT metric. Their Christoffel
symbols are
\begin{equation}
{{\Gamma}^{2}}_{33}=\frac{e^{-<v_{r}>(1-{\tau}_{0})t}}{2}sin{\theta}\label{55}
\end{equation}
\begin{equation}
{{\Gamma}^{3}}_{23}=\frac{e^{-<v_{r}>(1-{\tau}_{0}t}}{2}sin{\theta}\label{56}
\end{equation}
From these expressions one obtains the following geodesic equations
\begin{equation}
\frac{dv^{1}}{dt}=0\label{57}
\end{equation}
which implies $v^{r}=v^{1}=v^{0}=constant$. This result is indeed
important since, it shows that actually our hypothesis that the
random radial flow is constant is feasible, and can be derived from
a geodesic flow. To simplify the remaining equations one chooses a
torsion ${\tau}_{0}=\frac{1}{2}$ , which yields
\begin{equation}
\frac{dv^{2}}{dt}-sin{\theta}(v^{3})^{2}=0\label{58}
\end{equation}
and
\begin{equation}
\frac{dv^{3}}{dt}-e^{[1-\frac{1}{2}<v_{r}>t]}{v^{2}}{v^{3}}=0\label{59}
\end{equation}
These two last equations together yield
\begin{equation}
\frac{dv^{3}}{dv^{2}}=e^{-\frac{1}{2}<v_{r}>t}\frac{v^{2}}{v^{3}}\label{60}
\end{equation}
Some algebra yields the following result
\begin{equation}
\frac{v_{s}}{v^{\theta}}=e^{(1+cos{\theta})<v_{r}>t}\label{61}
\end{equation}
Thus the toroidal flow is proportional to the random flow and to the
Lyapunov exponential stretching. Note that at $t=0$ there is an
equipartition between the poloidal and toroidal flows while as time
evolves the Lyapunov exponential stretching makes the toroidal flow
velocity to increase without bounds with respect to the poloidal
flow. This is actually analogous to the dynamo action. Equation
(\ref{57}) also shows that one the Lyapunov exponents is constant
while the other is not and therefore the flow is a non-Anosov flow.

\newpage

\section{Conclusions} In this paper the importance of
investigation the twisted torus dynamo hyperbolic chaotic maps and
flows, has been stressed. The ideas developed here followed the path
of realistic dynamo maps obtained from the Gilbert's \cite{21}
shearing generalization of Arnold's cat map. Actually the
Tang-Boozer's \cite{5} idea of pilling up the Alfven twist of
magnetic flux tubes is fully developed geometrically by applying
their Lyapunov like metric and the Riemann geometrical Ricca's model
for the MFTs. Though Ricca has investigated only force-free MFTs, in
this paper the fast growth of magnetic fields in MFTs, induces a
dynamo action through the Lyapunov fast dynamo exponential
stretching. It is easy to show that from the self-induced equation
that the twisted MFTs, with zero Lyapunov exponential stretching
leads to non-fast dynamo action in the tubes, a result which is
nothing but the statement of anti-dynamo theorem of Vishik's
\cite{22}, which has been recently discussed in the context of the
MFTs \cite{23}. Computation of geodesic flows in Arnold's fast
kinematic dynamo and in twisted MFT, shows that in both case the
flow depends on the exponential stretching given by the Lyapunov
numbers. It is shown that at the beginning of time, there is an
equipartition between chaotic toroidal and poloidal flows, while as
time evolves an instability occurs due to the action of Lyapunov
exponential stretching. This result was obtained from the Euler
equations by Friedlander and Vishik \cite{24}.
\section{Acknowledgements}: \newline I am very much in  debt to Jean Luc Thiffeault for calling to my attention, many aspect of non-Anosov maps
and stretching in chaotic flows. R. Ricca and Dmitry Sokoloff for
helpful discussions on the subject of this paper. Financial supports
from Universidade do Estado do Rio de Janeiro (UERJ) and CNPq
(Brazilian Ministry of Science and Technology) are highly
appreciated.

\end{document}